\documentstyle[epsf,12pt]{article}

\def\ct#1{{\cal #1}}

\def\bc{\begin{center}}
\def\ec{\end{center}}
\def\be{\begin{equation}}
\def\ee{\end{equation}}

\newcommand{\nn}{\nonumber}

\newcommand{\RI}{\mbox{\scriptsize RI}}

\newcommand{\msbar}{\overline{\mbox{\scriptsize MS}}}
\newcommand{\MSbar}{\overline{\mbox{MS}}}
\newcommand{\bea}{\begin{eqnarray}}
\newcommand{\eea}{\end{eqnarray}}

\begin{document}
\pagestyle{empty} 
\vspace{-0.6in}
\begin{flushright}
Edinburgh 97/12 \\
FTUV/97-44 \\
IFIC/97-60 \\
ROME1-1180/97 \\
ROM2F/97/34

\end{flushright}
\vskip 0.2 cm
\centerline{\LARGE{\bf{Lattice $B$-parameters}}}
\centerline{\LARGE{\bf{for $\Delta S = 2$ and $\Delta I = 3/2$ Operators}}}
\vskip 0.3cm
\centerline{\bf{L. Conti$^a$, A. Donini$^b$, 
                V. Gimenez$^c$, G. Martinelli$^b$, 
                M. Talevi$^d$, A. Vladikas$^a$}}
\vskip 0.3cm
\centerline{$^a$ INFN, Sezione di Roma II, and 
Dip. di Fisica, Univ. di Roma ``Tor Vergata'',}
\centerline{Via della Ricerca Scientifica 1, I-00133 Roma, Italy.}
\smallskip
\centerline{$^b$ Dip. di Fisica, Univ. di Roma ``La Sapienza'' and
INFN, Sezione di Roma,}
\centerline{P.le A. Moro 2, I-00185 Roma, Italy.}
\smallskip
\centerline{$^c$ Dep. de Fisica Teorica and IFIC, Univ. de Valencia,}
\centerline{Dr. Moliner 50, E-46100, Burjassot, Valencia, Spain.}
\smallskip
\centerline{$^d$  Department of Physics \& Astronomy, University of Edinburgh,}
\centerline{The King's Buildings, Edinburgh EH9 3JZ, UK.}
\vskip 1.4cm

\abstract{ We compute several matrix elements of dimension-six 
four-fermion operators and extract their $B$-parameters. The calculations have 
been performed with the  tree-level Clover action at $\beta = 6.0$.
The renormalization  constants and mixing coefficients of the lattice operators
have been obtained non-perturbatively. 
In the $\MSbar$  renormalization scheme, at a renormalization scale 
$\mu \simeq 2$~GeV, we find $B_K \, (B_9^{3/2}) = 0.66(11), \, 
B_7^{3/2} = 0.72(5)$ and  $B_8^{3/2} = 1.03(3)$. 
The  result for $B_8^{3/2}$   has important 
implications for the calculation of $\epsilon^\prime / \epsilon$.}
\vfill\eject
\pagestyle{empty}\clearpage
\setcounter{page}{1}
\pagestyle{plain}
\newpage 
\pagestyle{plain} \setcounter{page}{1}

\section{Introduction}
\label{sec:intro}

The lattice evaluation of matrix elements of operators between hadronic states
is a necessary ingredient to the calculation, from first principles, of a wide
class of physical observables. In the effective Hamiltonian approach,
 weak amplitudes are expressed in terms of perturbative Wilson
coefficients multiplied by matrix elements of local operators, 
which can be evaluated on the lattice. The results are conventionally
presented in terms of $B$-parameters, which measure the deviation of the
matrix elements from their value in the Vacuum Saturation Approximation (VSA).
These quantities are subject to significant QCD corrections.

In this paper we focus on  $\Delta S = 2$ and $\Delta I = 3/2$ transition
amplitudes.
The former case  is characteristic of $K^0$--$\bar K^0$ oscillations,
which are related to indirect CP violation, parametrized by $\epsilon$.
We compute $B_K$, the  $B$-parameter of the matrix element
$\langle \bar K^0 \vert O^{\Delta S = 2} \vert K^0 \rangle$, where the
dimension-six, four-fermion operator $O^{\Delta S = 2}$
has a ``left-left" chiral structure (see sec.~\ref{sec:phen}). We also compute
the $\Delta I=3/2$ amplitudes
relevant in $K \rightarrow \pi\pi$ decays. These amplitudes are essential 
for theoretical predictions of direct CP-violation, parametrized by
$\epsilon^\prime$. Matrix elements of the form
$\langle \pi\pi \vert O^{3/2} \vert K \rangle$ enter in the calculation of
$\epsilon^\prime$, with two of the electro-penguin operators
($O^{3/2}_7$ and $O^{3/2}_8$) having a ``left-right" chiral structure, and one
operator ($O^{3/2}_9$) having a ``left-left" one.
Using Chiral Perturbation Theory (CPTh), 
these matrix elements  can be related to the 
single-state matrix elements $\langle \pi^+ \vert O^{3/2} \vert K^+ \rangle$.
In this work we compute the latter, parametrized in terms of  
$B$-parameters denoted  as $B_7^{3/2}$ and $B_8^{3/2}$ (the conventional basis
of operators for the $\Delta S=1$ effective Hamiltonian can be found 
in refs.~\cite{ds=1}; we also remind the reader that $B_9^{3/2}=B_K$).

The Wilson lattice regularization breaks chiral symmetry. This implies
that the $\Delta S = 2$ operator, which would otherwise renormalize
multiplicatively, mixes with operators belonging to different chiral
representations~\cite{MARTIW,octet}. The same is true for the two $\Delta I = 3/2$ 
operators, which would otherwise only mix with each other, as they belong to 
the same chiral representation.  Because of the mixing induced by the lattice,
the correct chiral behaviour of the operators is
achieved with Wilson fermions only in the continuum limit. For the $\Delta S
= 2$ case, for example, restoration of the chiral properties amounts
to the vanishing of the matrix element as $m_K
\to 0$. In practice, the mixing of the $\Delta S = 2$ operator with
operators of ``wrong" na\"\i ve chirality, computed at small but finite cutoff
$a^{-1}$, spoils the expected chiral behaviour \cite{BERNARD2,GAVELA}~\footnote{
In the Staggered fermion approach, where chiral symmetry is
partially preserved, the $\Delta S = 2$ matrix element displays the
correct chiral behaviour. Thus, the $B_K$-parameter obtained with staggered
fermions \cite{SHARPE} has been deemed more reliable.}. 
This problem can be attributed to two sources of systematic
error  in the computation of the matrix elements, namely 
the determination of the mixing coefficients in one-loop
perturbation theory, and the $\ct{O}(a)$ ($a$ is the lattice spacing)
discretization errors. Several methods have been proposed in order to improve
the determination of the mixing coefficients. One of them consists in evaluating
the renormalization constants computed in Standard Perturbation Theory (SPT)
using an effective coupling~\cite{LEPAGE} which should reduce higher order
corrections; we refer to it as Boosted Perturbation Theory
(BPT). Another is the Non-Perturbative Method (NPM) for the computation of the
renormalization constants on quark and gluon external states, as
proposed in ref.~\cite{NP}. Finally, in the spirit of ref.~\cite{octet},
the lattice mixing coefficients
can also be obtained non-perturbatively by using the Ward Identity Method (WI's), 
with external quark and gluon states~\cite{JAPBK}; see also refs.~\cite{WI}.
On the other hand,
recent studies \cite{DS=2}--\cite{wustl} 
(see also the present work) have shown that reducing the
discretization error by using the tree-level Clover action does not
improve  the chiral behaviour of the matrix elements, even when BPT is
implemented in the definition of the renormalized $\Delta S=2$
operator. Instead, a  good chiral
behaviour has been observed by evaluating the renormalization constants 
non-perturbatively, either with the NPM (refs. \cite{DS=2}--\cite{wustl} and
this work) or by using WI's (ref.~\cite{JAPBK}). The fact that the restoration
of the correct chiral behaviour has been seen both with the tree-level Clover
action (refs.~\cite{DS=2}--\cite{wustl} and the present work) and with the
Wilson action
(ref.~\cite{JAPBK}), suggests that discretization effects are less important 
than those due to the perturbative evaluation of the mixing coefficients.
Therefore, the recent Wilson fermion estimates of $B_K$ from the NPM or the
WI's, are considerably more reliable than those of earlier studies, based
on perturbative renormalization.

Given the success of the NPM  in the computation of $B_K$, we also apply 
it to the 
evaluation of the two $\Delta I=3/2$  $B$-parameters of the electro-penguin 
operators, $B_7^{3/2}$ and $B_8^{3/2}$. A recent calculation, using
the Wilson action and with the renormalization constants obtained in
BPT, found $B_8^{3/2} = 0.81(3)$~\cite{gupta_bp}, as opposed to the  earlier
results $B_8^{3/2} \simeq 1$ of refs.~\cite{busa,franco}. 
In the present work, using the Clover action and the NPM for the
evaluation of the renormalization constants we find $B_8^{3/2} = 1.03(3)$.
Notice that, although  obtained with a different lattice fermion action,
our BPT estimate  $B_8^{3/2} = 0.83(2)$ is, instead, fully compatible with
that of  ref.~\cite{gupta_bp}.  Our preferred value $B_8^{3/2} = 1.03(3)$,
obtained  with an improved operator renormalized non-perturbatively, only
suffers from discretization errors which are $\ct{O}(g_0^2 a)$. 
We thus believe that it  is  more reliable than previous results which have
been obtained with the (non-improved) operator renormalized in one-loop
perturbation theory. The same situation characterizes $B^{3/2}_7$: our NPM
and BPT results are not in agreement, but the latter is compatible with
the value quoted in \cite{gupta_bp}.
We stress that a precise determination of $B^{3/2}_8$,
combined with an equally reliable estimate of the strange quark mass, is
essential to the determination of the ratio $\epsilon^\prime/\epsilon$. For
example, the uncertainties in the measurement of $B^{3/2}_8$, combined with
the controversial results for $m_s$~\cite{gupta_ms}--\cite{chetyrkin}
 may change $\epsilon^\prime/\epsilon$ up to a factor of 2 
to 3. In view of its importance, we believe that $B^{3/2}_8$ should also
be computed by applying the NPM with the Wilson action also and/or by using
WI's for the determination of the lattice mixing coefficients of the
renormalized operator.

An extensive study of the renormalization properties of the four-fermion 
operators can be found in \cite{4ferm_teo}. There we detail all the 
theoretical and numerical issues of relevance to the non-perturbative 
renormalization of the $\Delta S = 2$ and $\Delta I=3/2$ operators. 
We have used these results in the present study.

The paper is organized as follows: in sec.~\ref{sec:phen} we introduce
the operators of interest and review briefly their phenomenological
implications; in sec.~\ref{sec:npm} we address the problem of operator
mixing and give a brief account of the NPM for the
computation of the renormalization constants; in sec.~\ref{sec:res} define
the $B$-parameters and discuss their extraction from lattice correlation
functions; in sec.~\ref{sec:numres}, we 
present our results for $B_K$, $B_7^{3/2}$ and $B_8^{3/2}$; finally, in
sec.~\ref{sec:concl} we compare our results to those previously obtained in
the literature.

\section{ $\Delta S = 2$ and
$\Delta I = 3/2$ transitions}
\label{sec:phen}

Schematically, the $\Delta S=2$ effective Hamiltonian has the form
\be
{\cal H}^{\Delta S = 2} = \frac{G^2_F}{16 \pi^2} M^2_W 
\hat O^{\Delta S=2}(\mu) \Phi(x_c,x_t,\mu)
\ee
where $x_q = m^2_q /M^2_W$ and $\Phi$ is a combination of the Inami-Lim 
functions \cite{INALIM} weighted by the CKM matrix elements. All perturbative 
QCD corrections (known at the next-to-leading order (NLO) \cite{BURAS}) 
are included in $\Phi$
(which is, hence, $\mu$-dependent). The rest of the notation is standard:
$G_F$ is the weak Fermi coupling, $M_W$ the mass of the $W$ boson and
$\mu$ the renormalization scale. The $\Delta S = 2$ operator in the
above expression is the renormalized one. It is defined as follows: 
\be
O^{\Delta S = 2} = \bar s \gamma^L_\mu d \bar s \gamma^L_\mu d ,
\label{eq:ods2}
\ee
where $\gamma^L_\mu = \gamma_\mu ( 1 - \gamma_5)$.
The  CP-violation parameter $\epsilon_K$ is defined as:
\be
\vert \epsilon_K \vert = \frac{G^2_F}{16 \pi^2} \frac{M^2_W}{\sqrt{2}
\Delta m_K} \left( \frac{8}{3} f^2_K m_K B_K(\mu) \right) \Phi(x_c,x_t,\mu),
\label{eq:eps}
\ee
where $\Delta m_K$ is the $K^0_L$-$K^0_S$ mass splitting. $m_K$ and $f_K$
denote the $K$-meson mass and decay constant, respectively.

In the $\Delta S = 1$ case, the $\Delta I=3/2$ 
contribution  to $\epsilon^\prime$ can be written as
\be
\label{eq:a_32}
{\rm Im} {\cal A}^{3/2} \propto - G_F {\rm Im} \lambda_t
\left [ C_7 \langle O_7^{3/2} \rangle_{VSA} B_7^{3/2} +
        C_8 \langle O_8^{3/2} \rangle_{VSA} B_8^{3/2} +
        C_9 \langle O_9^{3/2} \rangle_{VSA} B_9^{3/2} \right] ,
\ee
where $\lambda_t = V_{ts}^\ast V_{td}$ contains the CKM matrix dependence
and $\langle O^{3/2} \rangle_{VSA}$ stands for the VSA matrix element 
of the corresponding operator. The definitions of the operators can be found
for example in refs.~\cite{Ciuchini2,ban}. 
The Wilson coefficients, known to NLO~\cite{Ciuchini2}--\cite{Ciuchini}, 
are denoted by  $C_k\equiv C_k(M_W/\mu)$  and 
the operators $O_k$ ($k=7,8,9$) are defined as:
\bea
O_7^{3/2} &=& (\bar s_\alpha \gamma_\mu^L d_\alpha) 
\{ \bar u_\beta \gamma_\mu^R u_\beta - \bar d_\beta \gamma_\mu^R d_\beta \} 
  + (\bar s_\alpha \gamma_\mu^L u_\alpha)( \bar u_\beta \gamma_\mu^R d_\beta )
 , \nn \\
O_8^{3/2} &=& (\bar s_\alpha \gamma_\mu^L d_\beta) 
\{ \bar u_\beta \gamma_\mu^R u_\alpha - \bar d_\beta \gamma_\mu^R d_\alpha \} 
   + (\bar s_\alpha \gamma_\mu^L u_\beta) ( \bar u_\beta \gamma_\mu^R d_\alpha
 ), \label{q7_3/2} \\
O_9^{3/2} &=& (\bar s_\alpha \gamma_\mu^L d_\alpha) 
\{ \bar u_\beta \gamma_\mu^L u_\beta - \bar d_\beta \gamma_\mu^L d_\beta \} 
  + (\bar s_\alpha \gamma_\mu^L u_\alpha)( \bar u_\beta \gamma_\mu^L d_\beta )
, \nn
\eea
where $\gamma_\mu^R = \gamma_\mu (1 + \gamma_5)$ and
$\alpha,\beta = 1$--$3$ are colour indices.
The definitions of the $B$-parameters of eqs.~(\ref{eq:eps}) and
(\ref{eq:a_32}) will be given in sec.~\ref{sec:res}.

Two observations are necessary at this point.  Firstly,
since we are interested  in computing the matrix elements
$\langle \bar K^0  \vert \hat O^{\Delta S = 2} \vert K^0 \rangle$ and
$\langle \pi^+ \vert \hat O_k^{3/2} \vert K^+ \rangle$ (with $k = 7,8,9$),
only  the parity-conserving parts
of the operators of eqs.~(\ref{eq:ods2}) and (\ref{q7_3/2})  enter 
in the calculation.  Secondly, on the lattice the above matrix
elements are obtained in the standard way by studying the asymptotic behaviour,
at large time separations, of hadronic correlation functions of the form
$\langle P(y) \hat O(0) P(x) \rangle$ (see eqs.~(\ref{eq:corrs}) below), 
with $P$ denoting suitable pseudoscalar
densities which we use as meson sources and sinks. The Wick contractions of the
quark fields in the correlation functions
give rise to diagrams which are both ``eight"-shaped and ``eye"-shaped. 
The latter, however, cancel in the limit of degenerate up and down quark 
masses. Since our results are obtained in this limit, complicated subtractions 
of lower dimensional operators, necessary for the removal of the power 
divergences of the ``eye"-diagrams, are avoided.

\section{Non-Perturbative renormalization}
\label{sec:npm}

The NPM for the evaluation of the renormalization constants of lattice
operators consists in imposing suitable renormalization conditions
on lattice amputated quark correlation functions \cite{NP}.
In our case, we compute four-fermion Green functions in the Landau gauge.
All external quark lines are at equal momentum $p$. After amputating
and projecting these correlation functions (see refs.~\cite{DS=2} and
\cite{4ferm_teo} for details), the renormalization conditions are imposed in 
the deep Euclidean region at the scale $p^2 = \mu^2$. This 
renormalization scheme has been recently called the
 Regularization Independent (RI)  scheme~\cite{Ciuchini2} 
(MOM in the early literature) in order to emphasize that the renormalization
conditions are independent of the regularization scheme, although they depend 
on the external states used in the renormalization procedure (and on the
gauge). Thus, at fixed 
cutoff (i.e. fixed $\beta$), we compute non-perturbatively  the renormalization
constants and the renormalized 
operator $\hat O^{RI}(\mu)$ in the RI scheme. In order to obtain the physical
amplitudes, which are renormalization group invariant and scheme 
independent, the renormalized matrix 
elements must subsequently be combined with the corresponding Wilson 
coefficients of the  effective Hamiltonian. The latter are known 
in  continuum perturbation theory, at the
NLO,  both in the $\MSbar$  scheme~\cite{BURAS}--\cite{bjl}
and in the RI scheme~\cite{Ciuchini2}. Although not strictly necessary,
since the standard practice consists in
giving the $B$-parameters in the $\MSbar$ scheme, we will express
our results both in the $\MSbar$ and RI schemes.  
In order to obtain the corresponding operators in the $\MSbar$  scheme,
$\hat O^{\msbar}(\mu)$, the  matrix elements of the operators 
$\hat O^{RI}(\mu)$ must be corrected by  finite matching 
coefficients~\cite{Ciuchini2}.
We stress that, if $\mu$ is much larger than $\Lambda_{QCD}$, the NPM of
ref.~\cite{DS=2} and the WI method of
ref.~\cite{JAPBK}, used for the computation of the mixing coefficients of the
lattice operators, are equivalent, in the chiral limit. This has been shown for
two-fermion operators in ref.~\cite{NP}; for four-fermion operators it is
discussed in detail in refs.~\cite{4ferm_teo,testa} (see also
ref.~\cite{octet}).

In \cite{4ferm_teo}, we have determined non-perturbatively the operator mixing
for the complete basis of four-fermion operators, with the aid of the discrete
symmetries (parity, charge conjugation and switching of flavours),
in the spirit of ref.~\cite{BERNARD2}.
The renormalization of the parity-conserving operators, relevant to this work,
is conveniently expressed in terms of the following basis of five operators:
\bea
Q_1 &=& V \times V + A \times A , \nonumber \\
Q_2 &=& V \times V - A \times A , \nonumber \\
\label{base}
Q_3 &=& S \times S - P \times P , \\
Q_4 &=& S \times S + P \times P , \nonumber \\
Q_5 &=& T \times T .              \nonumber
\eea
The operators $Q_1,\dots,Q_5$ form a complete basis on the lattice.
In these expressions, $\Gamma \times \Gamma$ (with $\Gamma = V,A,S,P,T$ 
a generic Dirac matrix) stands for $\frac{1}{2}(
\bar \psi_1 \Gamma \psi_2 \bar \psi_3 \Gamma \psi_4 +
\bar \psi_1 \Gamma \psi_4 \bar \psi_3 \Gamma \psi_2) $, where 
$\psi_i,~i=1,\dots,4$ are fermion fields with flavours chosen
so as to reproduce the desired operators (see ref.~\cite{4ferm_teo} for
details). 
More specifically, the parity-conserving component of the four-fermion 
operator $O^{\Delta S = 2}$ corresponds to $Q_1$ in our basis. On the lattice,
this operator mixes under renormalization with the other four operators
as follows
\be
\label{eq:bk_sub}
\hat Q_1 = Z_{11} \left [ Q_1 + \sum^5_{i=2} Z_{1i} Q_i \right ], \nonumber
\ee
where $Z_{11}$ is a multiplicative logarithmically divergent renormalization
constant; it depends on the coupling and $a\mu$. The mixing coefficients 
$Z_{1i}$ (with $i = 2,\ldots,5$) are finite; they only depend on the lattice
coupling $g_0^2(a)$.

The renormalization of the parity-conserving parts of the
operators $O_7^{3/2}$ and $O_8^{3/2}$ is related to that of
the operators $Q_2$ and $Q_3$; the correspondence is given by
\bea
O_7^{3/2} & \rightarrow & Q_2 , \nonumber \\
O_8^{3/2} & \rightarrow & - 2 Q_3 . \nonumber 
\eea
The renormalized operators are defined as:
\begin{eqnarray}
\hat Q_2 &=& Z_{22} Q_2^s + Z_{23} Q_3^s , \nn \\
\hat Q_3 &=& Z_{32} Q_2^s + Z_{33} Q_3^s ,
\label{eq:q23ren}
\end{eqnarray}
where $Z_{ij}$ (with $i,j = 2, 3$) are logarithmically divergent
renormalization constants which depend on the coupling and $a\mu$.
The above mixing matrix is not peculiar to the lattice regularization, 
but also occurs in the continuum. The breaking of chiral symmetry by the 
Wilson action requires the additional subtractions:
\be
Q_i^s = Q_i + \sum_{j=1,4,5} Z_{ij} Q_j, \;\;\; i=2,3. \nn
\ee
where the $Z_{ij}$s are finite coefficients which only
depend on $g_0^2(a)$. 

Finally,  the operator $O^{3/2}_9$
corresponds to the operator $Q_1$ of eqs.~(\ref{base}). Thus, its 
renormalization constants, $B$-parameter, 
etc. are identical to those of $O^{\Delta S = 2}$.
The results for all the renormalization constants $Z_{ij}$ 
(computed with the NPM at several renormalization scales $\mu$ at $\beta = 
6.0$) can be found in \cite{4ferm_teo}.

\section{$B$-parameters}
\label{sec:res}

In the $\Delta S = 2$ case, the $B$-parameter is defined as
\be
B_K(\mu) = \frac{ \langle \bar K^0 \vert \hat O^{\Delta S = 2} (\mu) \vert
K^0 \rangle } {\langle \bar K^0 \vert \hat O^{\Delta S = 2} \vert
K^0 \rangle_{VSA} } \,\, ,
\label{eq:bkdef}
\ee
where the matrix element in the VSA is given by
\be
{\langle \bar K^0 \vert \hat O^{\Delta S = 2} \vert K^0 \rangle_{VSA} } = 
2 \left ( 1 +  \frac{1}{N_c} \right) Z_A^2 \vert \langle K^0 \vert A_\mu
\vert 0 \rangle \vert ^2 
= 2 \left ( 1 + \frac{1}{N_c} \right ) m_K^2 f_K^2 \,\, ,
\label{eq:ds2vsa}
\ee
with $Z_A$ the (finite) renormalization constant of the lattice axial current
$A_\mu = \bar \psi \gamma_\mu \gamma_5 \psi$. 
Since $\langle \bar K^0 \vert \hat O^{\Delta S = 2} \vert K^0 \rangle_{VSA}$
is given in terms of physical quantities ($m_K$ and $f_K$),
$B(\mu)$ runs with the scale $\mu$ exactly like the 
corresponding renormalized operator $\hat O(\mu)$. 
Another definition of the VSA matrix element has recently been proposed in
ref.~\cite{JAPBK}. It consists in vacuum-saturating each operator which enters
in the subtraction of eq.~(\ref{eq:bk_sub}). In this way, statistical 
fluctuations are reduced in the ratio of correlation functions
used to extract $B_K$ (defined as $R^{\Delta S=2}$ in eqs.~(\ref{eq:rats})
 below). 
As pointed out in ref.~\cite{Tassos}, however,
the definition used in ref.~\cite{JAPBK} spoils the good
scaling properties of $B_K(\mu)$. Thus, we insist on retaining
the standard definition of eq.~(\ref{eq:ds2vsa}), at the price of having
larger statistical errors.

For the $\Delta I = 3/2$ transitions, the $B$-parameters are defined
by:
\begin{eqnarray}
B_7^{3/2}(\mu) = \frac{ \langle \pi^+ \vert \hat O^{3/2}_7(\mu) \vert K^+
\rangle }{ \langle \pi^+ \vert \hat O^{3/2}_7 \vert K^+ \rangle_{VSA} } , \nn \\
B_8^{3/2}(\mu) =  \frac{ \langle \pi^+ \vert \hat O^{3/2}_8(\mu) \vert K^+
\rangle }{ \langle \pi^+\vert \hat O^{3/2}_8 \vert K^+ \rangle_{VSA} }.
\label{eq:b78def}
\end{eqnarray}
As discussed in sec.~\ref{sec:npm}, $B^{3/2}_9 (\mu) = B_K(\mu)$
in the limit of degenerate quark masses.
The VSAs for the above matrix elements depend on two different contributions 
of the form:
\bea
{\langle \pi^+ \vert \hat O^{3/2}_7 \vert K^+ \rangle_{VSA} }
&=& \frac{2}{N_c} Z^2_P  \langle \pi^+ \vert P \vert 0 \rangle \langle 0
\vert P \vert K^+ \rangle - Z_A^2 \langle \pi^+ \vert A_\mu \vert 0 \rangle
\langle 0 \vert A_\mu \vert K^+ \rangle, \nn
\\
{\langle \pi^+ \vert \hat O^{3/2}_8 \vert K^+ \rangle_{VSA} } &=& 2 Z^2_P  
\langle \pi^+ \vert P \vert 0 \rangle \langle 0 \vert P \vert K^+ \rangle 
- \frac{Z_A^2}{N_c} \langle \pi^+ \vert A_\mu \vert 0 \rangle \langle 0 \vert
A_\mu \vert K^+ \rangle,
\label{eq:b78vsa}
\eea
where $Z_P$ is the renormalization constant of the lattice pseudoscalar
density $P=\bar \psi \gamma_5 \psi$ (renormalized at the same scale $\mu$
in the RI scheme). Since we work with degenerate quark masses, we have left
the flavour content of the operators $A_\mu$ and $P$ unspecified; they
are meant to have whatever flavour is required by the hadronic states of
their matrix elements (i.e. $P_\pi$, $P_K$ and similarly for $A_\mu$).
Contrary to the $\Delta S=2$ case,
the leading terms of the above VSA matrix elements are $\mu$-dependent
quantities, with an anomalous dimension equal to
twice the anomalous dimension of the pseudoscalar density $P$.
Thus the $B$-parameters do not scale in $\mu$ like the matrix elements
of the corresponding operators. The last terms on the r.h.s. of
eqs.~(\ref{eq:b78vsa}) vanish in the chiral limit. Consequently,
since  we are ultimately interested in passing from the $\langle \pi^+ \vert 
\hat O^{3/2} \vert K^+ \rangle$
matrix elements to the $\langle \pi \pi  \vert \hat O^{3/2} \vert K \rangle$
ones using soft pion theorems, following ref.~\cite{franco}, we have dropped
the last terms on the r.h.s. of eqs.~(\ref{eq:b78vsa}).
In order to extract the $B$-parameters, 
we need to compute the following two- and three-point correlation
functions:
\bea
G_P(t_x,\vec p) = \sum_{\vec x} 
\langle P(x) P(0) \rangle e^{-\vec p \cdot \vec x} &,&
\,\,\,\,\,\,\,\,\,\,\,
G_A(t_x,\vec p) = \sum_{\vec x} \langle A_0(x) P(0) \rangle e^{-\vec p \cdot \vec x},
\label{eq:corrs} \\
G_{\hat O}(t_x,t_y;\vec p, \vec q) &=& \sum_{\vec x,
\vec y} \langle P(y) \hat O(0) P(x)\rangle
e^{-\vec p \cdot \vec y} e^{\vec q \cdot \vec x} , \nn
\eea
where $x \equiv (\vec x, t_x), y \equiv (\vec y , t_y)$ and
$\hat O$ stands for any four-fermion operator of interest. As stated
above, all correlation functions have been evaluated with degenerate quark
masses and therefore only the ``eight-diagrams" contribute to $G_{\hat O}$.
By forming suitable ratios of the above correlations, and looking at their
asymptotic behaviour at large time separations, we isolate the desired
operator matrix elements. In particular the ratios:
\bea
 R^{\Delta S = 2} &=& \frac{1}{Z^2_A} \frac{G_{\hat O^{\Delta S=2}}}
{G_P G_P} \to \frac{\langle \bar K^0(\vec p) \vert {\hat O^{\Delta S=2}}
\vert K^0(\vec p) \rangle}{Z^2_A \vert \langle 0 \vert P \vert K^0 \rangle
\vert ^2 }, \nn \\
 R^{3/2}_7 &=& -\frac{N_c}{2 Z^2_P} \frac{G_{\hat O_7^{3/2}}}{G_P G_P} 
\to \frac{N_c}{2} \frac{ \langle \pi^+ \vert {\hat O_7^{3/2}} \vert
K^+ \rangle} {Z^2_P \langle \pi^+ \vert P \vert 0 \rangle 
\langle 0 \vert P \vert  K^+ \rangle } , \label{eq:rats} \\
 R^{3/2}_8 &=& -\frac{1}{2Z^2_P} \frac{G_{\hat O_8^{3/2}}}{G_P G_P} \to
\frac{1}{2} \frac{ \langle \pi^+ \vert {\hat O_8^{3/2}} \vert K^+ \rangle }
{Z^2_P \langle \pi^+ \vert P \vert 0 \rangle 
\langle 0 \vert P \vert  K^+ \rangle } , \nn
\eea
give, up to computable factors, the $B$-parameters of interest. For
comparison, we have obtained results with the operator renormalized 
not only with the NPM, but also with SPT and BPT.

\section{Numerical results}
\label{sec:numres}
Our simulation has been performed at $\beta = 6.0$ with the tree-level Clover
action in the quenched approximation. Quark masses have been fixed at
$\kappa = 0.1440, 0.1432$ and $0.1425$. The renormalization constants have
been obtained from quark correlation functions, in the Landau gauge,
on a $16^3 \times 32$ lattice, with  100 configurations. The hadronic matrix
elements have been computed on an $18^3 \times 64$ lattice with $460$
configurations. Details on the choice of time intervals, 
spatial momenta and related
technicalities are to be found in ref.~\cite{B_K}.
Statistical errors have been estimated with the jacknife method, by decimating
46 configurations at a time. We have neglected the statistical errors
of the renormalization constants, quoting only those of the hadronic matrix 
elements. 
In the above ratios, we also need the (finite) axial-current  
renormalization constant $Z_A$
and the $a\mu$-dependent renormalization constant $Z_P$
of the pseudoscalar density. Depending on the method of renormalization
of the four-fermion operator (NPM, SPT or BPT), we have used the corresponding
estimate of $Z_A$ and $Z_P$, obtained with the same method of calculation.
Although $Z_A$ should not depend~\footnote{For the tree-level Clover 
action, the best available non-perturbative estimate for $Z_A$ is obtained 
from lattice axial WI's; at $\beta=6.0$ this is $Z_A = 1.11(2)$ \cite{clv}.}
on $a\mu$, slight variations of its NPM estimate,  arising 
from systematic effects, partially
cancel analogous variations of  $R^{\Delta S=2}$,
giving much more stable results. The NPM estimates for $Z_P$ and $Z_A$
used in the present work are those of ref.~\cite{ggrt}.

In order to extract the $B$-parameters from the ratios of
eqs.~(\ref{eq:rats}), we follow the procedure of ref.~\cite{B_K}, fitting
them with the function
\be
\label{eq:r_1}
R = \alpha + \beta X + \gamma Y \, ,
\ee
where
\bea
X = \frac{8}{3} \frac{G_A G^\dagger_A}{G_P G_P} \to \frac{8}{3}
\frac{f_K^2 m_K^2}{Z_A^2 \vert \langle 0 \vert P \vert K^0 \rangle \vert ^2}
\, , \,\,\,\,\,\,\,\,\,\,\,
Y = \frac{( p \cdot q)}{m^2_K} X 
\label{fit_var}
\eea
(the large time asymptotic behaviour of $X$ is also shown above).
The results of the fits are collected in tab.~\ref{tab2}, 
at eight different values of the  renormalization scale $\mu^2 a^2$.
We also show results obtained with the renormalization
constants evaluated in SPT and in BPT. In the latter case, we use
the recipe $\alpha_s^{boost} = \alpha_s / (\frac{1}{3} Tr \langle U_\Box 
\rangle )$, where $\langle U_\Box \rangle$ stands for the average plaquette. 
At $\beta = 6.0$, we have used $\alpha_s^{boost} = 1.68/4\pi$.
{\scriptsize
\begin{table}[t]
\centering
\begin{tabular}{|c|c|c|c|c|c|}
\hline
Operator & $\mu^2 a^2$ & $\alpha$ & $\beta$& $\gamma$ \\
\hline \hline  
&$0.31$ & $ 0.027 \pm 0.014 $ & $ 0.17 \pm 0.16 $ & $ 0.68 \pm 0.12 $ \\
&$0.62$ & $-0.018 \pm 0.014 $ & $ 0.28 \pm 0.17 $ & $ 0.67 \pm 0.12 $ \\
&$0.96$ & $-0.014 \pm 0.014 $ & $ 0.27 \pm 0.16 $ & $ 0.66 \pm 0.11 $ \\
&$1.27$ & $-0.010 \pm 0.013 $ & $ 0.26 \pm 0.16 $ & $ 0.66 \pm 0.11 $ \\
&$1.39$ & $-0.004 \pm 0.013 $ & $ 0.23 \pm 0.16 $ & $ 0.66 \pm 0.11 $ \\
$R^{\Delta S = 2}$
&$1.85$ & $-0.005 \pm 0.013 $ & $ 0.25 \pm 0.16 $ & $ 0.66 \pm 0.12 $ \\
&$2.46$ & $ 0.002 \pm 0.013 $ & $ 0.25 \pm 0.16 $ & $ 0.67 \pm 0.12 $ \\
&$4.01$ & $ 0.012 \pm 0.012 $ & $ 0.25 \pm 0.16 $ & $ 0.68 \pm 0.12 $ \\
& SPT    & $-0.069 \pm 0.013 $ & $ 0.17 \pm 0.16 $ & $ 0.65 \pm 0.12 $ \\
& BPT    & $-0.058 \pm 0.013 $ & $ 0.18 \pm 0.16 $ & $ 0.66 \pm 0.12 $ \\
\hline\hline
&$0.31$ & $1.70 \pm 0.16 $ & $ -6.4 \pm 1.3  $ & $ 6.3  \pm 0.7  $ \\
&$0.62$ & $0.81 \pm 0.07 $ & $ -0.8 \pm 0.6  $ & $ 3.5  \pm 0.4  $ \\
&$0.96$ & $0.69 \pm 0.04 $ & $ -0.1 \pm 0.4  $ & $ 2.6  \pm 0.3  $ \\
&$1.27$ & $0.70 \pm 0.03 $ & $  0.1 \pm 0.4  $ & $ 2.3  \pm 0.3  $ \\
$R^{3/2}_7$
&$1.39$ & $0.70 \pm 0.03 $ & $ 0.04 \pm 0.34 $ & $ 2.2  \pm 0.3  $ \\
&$1.85$ & $0.69 \pm 0.03 $ & $ 0.15 \pm 0.28 $ & $ 1.94 \pm 0.23 $ \\
&$2.47$ & $0.69 \pm 0.02 $ & $ 0.22 \pm 0.25 $ & $ 1.77 \pm 0.21 $ \\
&$4.01$ & $0.73 \pm 0.02 $ & $ 0.31 \pm 0.22 $ & $ 1.62 \pm 0.20 $ \\
& SPT   & $0.68 \pm 0.02 $ & $ 0.45 \pm 0.17 $ & $ 1.14 \pm 0.14 $ \\
& BPT   & $0.48 \pm 0.02 $ & $ 0.41 \pm 0.22 $ & $ 1.58 \pm 0.17 $ \\
\hline\hline
&$0.31$ & $ 1.20  \pm 0.06 $ & $ -1.0 \pm 0.4  $ & $ 0.66 \pm 0.20 $\\
&$0.62$ & $ 1.08  \pm 0.04 $ & $ -0.5 \pm 0.3  $ & $ 0.56 \pm 0.15 $\\
&$0.96$ & $ 1.04  \pm 0.03 $ & $ -0.3 \pm 0.3  $ & $ 0.53 \pm 0.14 $\\
&$1.27$ & $ 1.00  \pm 0.03 $ & $-0.06 \pm 0.25 $ & $ 0.52 \pm 0.13 $\\
$R_8^{3/2}$
&$1.39$ & $ 1.00  \pm 0.03 $ & $-0.05 \pm 0.25 $ & $ 0.52 \pm 0.13 $\\
&$1.85$ & $ 0.99  \pm 0.03 $ & $ 0.04 \pm 0.24 $ & $ 0.51 \pm 0.13 $\\
&$2.46$ & $ 0.99  \pm 0.02 $ & $ 0.09 \pm 0.24 $ & $ 0.51 \pm 0.13 $\\
&$4.01$ & $ 0.98  \pm 0.02 $ & $ 0.15 \pm 0.23 $ & $ 0.51 \pm 0.13 $\\
& SPT   & $ 0.81  \pm 0.02 $ & $ 0.36 \pm 0.17 $ & $ 0.42 \pm 0.09 $\\
& BPT   & $ 0.75  \pm 0.02 $ & $ 0.29 \pm 0.16 $ & $ 0.39 \pm 0.09 $\\
\hline
\end{tabular}
\caption{\it{ Values of the fit parameters for $R^{\Delta S = 2},
R^{3/2}_7$ and $R^{3/2}_8$, from the  NPM (at several renormalization scales), 
SPT and BPT.}} 
\label{tab2}
\end{table}
}

We first examine the results for $R^{\Delta S =2}$, from which $B_K$ can be
extracted.
As in \cite{DS=2}-\cite{wustl}, it can be seen that $\alpha$ and $\beta$
(which are lattice artifacts) are compatible with zero within at most
$2\sigma$, unlike their SPT and BPT values.
Thus (within our statistical accuracy and up to terms of ${\cal O}(g^2 a)$),
the correct chiral behaviour of the matrix elements is restored using the NPM.
We point out the stability of $\gamma$ as a function of $\mu$.
Assuming $\alpha$ and $\beta$ to be zero, the $B_K$-parameter is then given by:
\be
B_K = \gamma . \nn
\ee

In our original analysis \cite{DS=2,B_K},
we have erroneously considered  the mixing of the 
$\Delta S = 2$ operator with only the three other operators
which appear in one-loop 
perturbation theory \cite{MARTIW}. 
It has been stressed in \cite{JAPBK,wustl,gupta_bp,4ferm_teo}
that, non-perturbatively, $O^{\Delta S = 2}$ also mixes with a fourth operator.
The results of ref.~\cite{wustl} and the present work take into account
the complete non-perturbative mixing. 

We now turn to $R^{3/2}_7, R^{3/2}_8$, from which $B^{3/2}_7$
and $B^{3/2}_8$ can be extracted. Note the stability of $\alpha$ as
a function of $a\mu$ and the compatibility of $\beta$ with zero. In this 
case, for both  operators, the $B$-parameters are simply given by
\be
B^{3/2} = \alpha .
\ee
Comparing the NPM results to those obtained in SPT and BPT, we find agreement
between the NPM and SPT values of $B^{3/2}_7$, whereas its BPT value is
incompatible with the other two. The $B^{3/2}_8$ result depends on the method
used for its  renormalization.

\section{Physics results and conclusions}
\label{sec:concl}

The $B$-parameters which can be extracted from the results given in table
\ref{tab2} have been obtained in the RI scheme, since the
operators have been renormalized with the NPM.
\par As previously stated, it
is customary to express all results in the $\MSbar$ scheme (with NDR
dimensional regularization). The finite matching between the RI and
$\MSbar$-NDR renormalization schemes can be done in continuum 
perturbation theory, by computing the operator matrix element in the same 
gauge and on the same external quark states as those used for the 
non-perturbative calculation of the lattice renormalization constants.
At NLO this matching has been obtained in \cite{Ciuchini2}. We have used it
in order to convert our NPM results from the RI to the $\MSbar$-NDR
scheme. The SPT and BPT cases are less straightforward: we have started from
the renormalization constants of ref.~\cite{frez}, which relate
the lattice operators to those renormalized in the  $\MSbar$-DRED
scheme. To these constants
we have  subtracted  the contributions due to the use of
 the  DRED  scheme and added those  corresponding to  the RI one.
In this way we have obtained, for both  the
SPT and BPT cases, the renormalization constants in the RI scheme,
 from which the results of tab.~\ref{tab2} are computed. Alternatively, we have
also calculated and added the necessary $\MSbar$-NDR contributions in order to
obtain the SPT and BPT results in this scheme. The latter calculation
has also been performed in \cite{gupta_bp} for the Wilson action.
At NLO \cite{Ciuchini2}, the result for $B_K$ is
\be
B_K^{\msbar}(\mu)=
\left(1 - \frac{\alpha_s(\mu)}{4\pi}\Delta r_+^{\msbar}\right)
B_K^{\RI}(\mu) , \label{eq:mbk}
\ee
where 
\be 
\Delta r_+^{\msbar}=14/3-8\ln 2 . \nn
\ee
The matching relation for the renormalized operators $\hat O^{3/2}_i$ ($i=7,8$)
at NLO is also given in \cite{Ciuchini2}:
\be
\left ( \hat O_i^{3/2} \right )_{\msbar} =
\left ( \delta_{ij} - \frac{\alpha_s(\mu)}{4\pi} \Delta r_{ij}^{\msbar} \right )
\left ( \hat O_j^{3/2} \right )_{\RI}
\ee
where
\be
\Delta r_{ij}^{\msbar} = \left (
\begin{array}{cc} 
\frac{2}{3} + \frac{2}{3} \ln 2 & -2 -2 \ln 2 \\
2 - 2 \ln 2 & - \frac{34}{3} + \frac{2}{3} \ln 2 
\end{array}
\right ) . \nn
\ee
Hence, for the $B$-parameters $B_7^{3/2}$ and $B_8^{3/2}$ we have:
\be
\left ( 
\begin{array}{c}
B_7^{3/2} \\ B_8^{3/2} 
\end{array}
\right )_{\msbar} =
\frac{1}{ \left (1 + \frac{\alpha_s(\mu)}{4 \pi} \Delta r_P^{\msbar} 
\right )^2 }
\left ( 
\begin{array}{cc}
        1 - \frac{\alpha_s(\mu)}{4\pi} \Delta r_{77}^{\msbar} & 
         -N_c \frac{\alpha_s(\mu)}{4\pi} \Delta r_{78}^{\msbar} \\
 -\frac{1}{N_c} \frac{\alpha_s(\mu)}{4\pi} \Delta r_{87}^{\msbar} &
        1 - \frac{\alpha_s(\mu)}{4\pi} \Delta r_{88}^{\msbar}
\end{array}
\right )
\left ( 
\begin{array}{c}
B_7^{3/2} \\ B_8^{3/2} 
\end{array}
\right )_{\RI}
\label{eq:match}
\ee
with
\be 
\Delta r_P^{\msbar}= 16/3 . \nn
\ee
The uncertainty due to the choice of $\alpha_s(\mu)$ is about $\pm 0.03$.
We find a further uncertainty
of $\pm 0.03$ when varying the number of active flavours from $0$ to $4$.

In order to compare our result for $B_K$ with other theoretical predictions
of the same quantity, it is useful to convert it to the RGI quantity,
$\hat B_K$, by multiplying it by the Wilson coefficient.
At NLO, $\hat B_K$ is given by
\begin{equation}
\label{rgibp}
{\hat B}_K = \alpha_s (\mu)^{-\gamma^{(0)}/2 \beta_0} 
\left[ 1- \frac{\alpha_s (\mu)}{4\pi}
\left(\frac{\gamma^{(1)} \beta_0-\gamma^{(0)} \beta_1}{2\beta^2_0}\right)
\right] B^{\msbar}_K(\mu)\, , 
\end{equation}
where $\beta_{0,1}$ and $\gamma^{(0,1)}$ are 
the leading and next-to-leading coefficients of the $\beta$-function and
anomalous dimension.  $\beta_{0,1}$ and $\gamma^{(0)}$ are universal 
whereas $\gamma^{(1)}$ depends on the regularization and
renormalization schemes. The explicit expressions of these quantities 
can be found for example in \cite{BURAS}. This estimate
of $\hat B_K$ is also regularization-scheme independent, up to
next-to-NLO order terms. 

In tab. \ref{fine} we collect all our results for the $B$-parameters,
obtained by computing their renormalization with the NPM and in BPT,
at the reference scale $\mu \simeq 2$~GeV in the $\MSbar$ scheme.
For $B_K$ we also give the RGI value $\hat B_K$.
Moreover, we compare our results to the most
recent ones obtained with the Wilson action at the same coupling and with
operator renormalizations carried out perturbatively in BPT \cite{gupta_bp}.
\par
Two  points concerning this comparison deserve some attention here.
The first is that in ref.~\cite{gupta_bp} the results were obtained
in 1-loop BPT by renormalizing the lattice operators directly in the
$\MSbar$-NDR scheme, rather than doing the renormalization in the RI
scheme and then matching to $\MSbar$-NDR with eq.~(\ref{eq:match}), as
we have to do with the NPM.
In perturbation theory,
the two procedures differ by $\ct{O}(\alpha_s^2)$ terms, which
introduce a systematic difference  of about $12\%$--$15\%$
in the results for $ B_7^{3/2}$ and $B_8^{3/2}$. For this reason,
in order to make  the
comparison meaningful, our BPT estimates in the $\MSbar$-NDR
scheme, given  in Table~\ref{fine}, were obtained using the same procedure
as in \cite{gupta_bp}. The second  point is that in ref.~\cite{gupta_bp}, a more
complicated version of BPT than in this work has been implemented, 
which involves tadpole
resummations and a choice of ``optimal" scale $q^\ast$ for the boosted
coupling. Since the recipe for $q^\ast$ is not unique,
results with several choices of $q^\ast$ have been listed in \cite{gupta_bp},
and can be compared with our results. Finally, we point out that
a direct comparison of our $B_K$ value to those of \cite{JAPBK} 
is not possible, because the latter have been obtained at different
lattice couplings.
{\scriptsize \begin{table}[hptb]
\centering
\begin{tabular}{|c|c|c|c|}\hline
Quantity & Method & Result & Ref. \\
\hline
\hline
            & NPM                & 0.66(11) & this work       \\
$B_K$       & BPT                & 0.65(11) & this work       \\
            & BPT $q^\ast = 1/a$ & 0.74(4)  & \cite{gupta_bp} \\
\hline
            & NPM                & 0.93(16) & this work \\
$\hat B_K$  & BPT                & 0.92(16) & this work \\
\hline
            & NPM                  & 0.72(5)  & this work       \\
$B_7^{3/2}$ & BPT                  & 0.58(2)  & this work       \\
            & BPT $q^\ast = 1/a$   & 0.58(2)  & \cite{gupta_bp} \\
            & BPT $q^\ast = \pi/a$ & 0.65(2)  & \cite{gupta_bp} \\
\hline
            & NPM                  & 1.03(3)  & this work       \\
$B_8^{3/2}$ & BPT                  & 0.83(2)  & this work       \\
            & BPT $q^\ast = 1/a$   & 0.81(3)  & \cite{gupta_bp} \\
            & BPT $q^\ast = \pi/a$ & 0.84(3)  & \cite{gupta_bp} \\
\hline
\end{tabular}
\caption{\it{B-parameters for $\Delta S = 2$ and $\Delta I = 3/2$ operators 
at the renormalization scale $\mu = a^{-1} \simeq 2$~GeV. All 
results are in the $\MSbar$ scheme.}}
\label{fine}
\end{table}
}

Our results from NPM and BPT for $B_K$ and $\hat B_K$ are in perfect 
agreement. With a larger statistical error, our $B_K$ value also 
agrees with those of refs.~\cite{gupta_bp,blum}.
We find, instead, a discrepancy between our NPM and BPT estimates of
$B_7^{3/2}$ and $B_8^{3/2}$. Our values obtained with BPT are fully compatible
to those of \cite{gupta_bp} (at least for one $q^\ast$ value),
where the Wilson action was used. The NPM estimate, instead,  is
in disagreement with any value obtained in BPT (either with the Wilson or the
Clover action and for several boosting variants). This indicates that the
difference between our NPM estimate and that of ref.~\cite{gupta_bp} is due
to the NPM used in the former result, rather than the implementation
of different actions (Clover and Wilson respectively).
The increase in the value of $B_8^{3/2}$, obtained with the NPM, is of
great phenomenological interest, since it may induce a significant decrease
of the ratio $\epsilon^\prime/\epsilon$.

In conclusion,
the non-perturbative renormalization of the four-fermion operators,
either with the NPM or with WI's, strongly improves the reliability of lattice
computations of $B$-parameters with Wilson-like fermions. Perturbative mixing
of lattice operators, carried out at lowest order, fails to reproduce
the expected  chiral
behaviour of the matrix elements (at present day couplings). A good 
chiral behaviour of $O^{\Delta S=2}$ is restored when the operators are 
renormalized non-perturbatively. Moreover, $B^{3/2}_7$ and $B^{3/2}_8$
obtained with the NPM disagree by as much as $20\%$ from the ones obtained in
BPT.

\section*{Acknowledgements}
L.C., G.M. and A.V. acknowledge partial support by the M.U.R.S.T.; the same
authors and V.G. acknowledge partial support by the EU (contract no.
CHRX-CT93-0132). V.G. acknowledges the partial support by CICYT under 
grant number AEN-96/1718.   
M.T. acknowledges the PPARC for support through grant no. GR/L22744
and the INFN for partial support.

\end{document}